\documentclass[aps,nofootinbib,manuscript, appendix]{revtex4}
\usepackage{bm}
\usepackage{amssymb}
\usepackage{lscape}
\usepackage{amsmath}
\usepackage{graphicx}
\usepackage{epsfig}
\usepackage{appendix}
\begin{document}

\title{Geometrization of gravito-electromagnetic interactions from boundary conditions in the variational principle}
\author{ $^{2}$ Jes\'us Mart\'{\i}n Romero\footnote{E-mail address: jesusromero@conicet.gov.ar}, $^{2}$ Luis Santiago Ridao\footnote{E-mail address: santiagoridao@hotmail.com}, $^{1,2}$ Mauricio Bellini\footnote{E-mail: mbellini@mdp.edu.ar}}

\address{$^{1}$ Departamento de F\'{\i}sica, Facultad de Ciencias Exactas y
Naturales, Universidad Nacional de Mar del Plata, Funes 3350,
(7600) Mar del Plata,
Argentina.\\
$^2$ Instituto de Investigaciones F\'{\i}sicas de Mar del Plata (IFIMAR). Consejo Nacional de Investigaciones Cient\'{\i}ficas y T\'ecnicas (CONICET). }

\begin{abstract}
We study the conditions of integrability when the boundary terms are considered in the variation of the geometric contribution of the  Einstein-Hilbert action. We explore the emergent physical dynamics that is obtained when we make a displacement from a background Riemann manifold to an extended one, on which the non-metricity is nonzero. Under these circumstances, a classical description of electrodynamics and non-perturbative gravitational waves are considered in the extended manifold, when we variate the action.
\end{abstract}


\maketitle
\section{Introduction and motivation}\label{secintro}

Geometrodynamics \cite{Wh,BB7} is a picture of general relativity that studies the evolution of the spacetime geometry. The key notion of the geometrodynamics was the idea of {\it charge without charge}, in which the Maxwell field was taken to be free of sources, and hence a non-vanishing charge could only arise from an electric flux line trapped in the topology of spacetime. With the construction of ungauged supergravity theories, it was realised that the Abelian gauge fields in such theories were source-free, and so the charges arising therein were therefore central charges\cite{G1}. The significant advantages of geometrodynamics, usually come at the expense of manifest local Lorentz symmetry \cite{R}. During the 70s and 80s decades a method of quantization was developed in order to deal with some unresolved problems of quantum field theory in curved spacetimes \cite{pru,pru2,pru3}. Quantum geometrodynamics, introduced by Wheeler  \cite{BB6,BB7}, and quantum geometry \cite{BB4,BB5} are some of the geometrical frameworks involved.
Recently, a quantum Unified Spinor Field (USF) was developed (see \cite{arrb} and references therein), where a quantum spinor field with components $\hat{\Psi}_{\alpha}$, is responsible for the displacement from the Riemannian manifold to an extended one. These fields are defined on the background spacetime, emerging from the expectation value of the quantum structure of spacetime generated by matrices that comply with a Clifford algebra. In this framework, was demonstrated that spinor fields are candidate to describe all known interactions in physics, with gravitation included, and a non-linear description of Gravitational Waves (GW) can be done. However, in this work we shall refer
only to classical physics. The boundary conditions in the minimum action's principle is a very important issue which must be taken into account to develop a physical theory\cite{york,gh}. In this framework, the conditions of integrability of the boundary conditions are relevant. In this paper we are aimed to study the possible electromagnetic physics which can be obtained when we consider an displacement from a background Riemannian manifold, such that the non-metricity on the extended manifold is nonzero.

The work is organized as follows: In Sect. (\ref{secdyn}) study the geometrical origin of sources in the geodesic equations which can be related to nontrivial boundary conditions in the minimal action principle. In Sect. (\ref{eins}), we show that the Einstein equations with cosmological parameter included in absence of physical sources on a Riamannian manifold are equivalent to these equations with sources, but with a different cosmological parameter, when the extended manifold is displaced with respect to the Riemann one by contortional terms. In Sect. (\ref{secwave}), we study massless gravitational waves equations in this context. In Sect. (\ref{ge}) we study a particular case where the contortion term, which characterizes the extended manifold, is originated by electromagnetic fields. Finally, in Sect. (\ref{fc}) we develop some final remarks.

\section{Boundary conditions in the geometrical action}\label{secdyn}

In a previous paper we have studied the dynamics obtained from a variational principle over the Einstein-Hilbert (EH) action, where non-metricity\cite{grav} is considered. In this work, We shall deal with an extended geometry, by considering only the geometric part of the EH action
\begin{eqnarray}\label{a1}
{\cal S}\,=\,\frac{1}{2\kappa}\, \int_{V}d^{4}x\,\sqrt{-g}\,{\cal R},
\end{eqnarray}
in which $\cal R$ is the scalar curvature of the complete (non-Riemannian) connection. The variational principle is expressed as $\delta {\cal{S}}=0$, with \begin{eqnarray}\label{a1b}
\delta{\cal S}\,=\,\frac{1}{2\kappa}\, \int_{V}d^{4}x\,\sqrt{-g}\left(\delta\,g^{\alpha\beta}\,G_{\alpha\beta}\,+\,g^{\alpha\beta}\,\delta{\bar{R}_{\alpha\beta}}\,+\,g^{\alpha\beta}\,\delta\Delta R_{\alpha\beta}\right),
\end{eqnarray}
in which
\begin{equation}\label{ac1}
R_{\alpha\beta}\,=\,\bar{R}_{\alpha\beta}\,+\, \Delta\, R_{\alpha\beta} ,
\end{equation}
$\bar{\cal R}$ being the Riemannian scalar curvature and $\Delta \cal R$ is the part of $\cal R$, which is exclusively non-Riemannian. The interesting here is that, it is obtained new physics from last two terms of the r.h.s., of eq. (\ref{a1b}). We shall refer to this issue later.\\

Now we consider the equation (\ref{a1b}) in presence of non-metricity. The variation of the action (\ref{a1}), is
\begin{eqnarray}\label{1.8din}
\delta {\cal S}\,=\,\int_{V} d^{4}x\, \sqrt{-g} \, G_{\alpha\beta}\delta
g^{\alpha\beta} \,+\,\int_{V}d^{4}x\, \sqrt{-g}\,
N_{\mu\alpha\beta}\,W^{\beta\mu\alpha},
\end{eqnarray}
in which auxiliary tensor $W^{\mu}_{\alpha\beta}$ is defined through the variation of connections $\Gamma^{\mu}_{\alpha\beta}$, by
\begin{equation} \label{a12}
W^{\mu}_{\alpha\beta}\,=\,\delta\Gamma^{\mu}_{\alpha\beta}\,-\,\delta\Gamma^{\sigma}_{\sigma\beta}\,\delta^{\mu}_{\alpha}.
\end{equation}
An extra term must be introduced in eq. (\ref{1.8din}), in order to take into account the torsion effects. However, in this work those contributions will not be considered. Furthermore, we shall neglect the surface term in the equation (\ref{1.8din}). This one could be interpreted as a source of the cosmological parameter in a gaussian hypersurface. The surface term is obtained in virtue of the Stokes theorem.
When the condition of integrability is fulfilled, the boundary terms in (\ref{1.8din}), takes the form
\begin{eqnarray}\label{super}
\int_{ V}d^{4}x \,\sqrt{-g}\,
\left(W^{\mu}\right)_{|\mu}\,=\,\int_{\partial V}d^{3}x\, \sqrt{-g}\,
W^{\mu}n_{\mu},
\end{eqnarray}
where $n_{\mu}$ is a vector field which is normal to the 3D
hypersurface $\partial V$, involving the manifold of 4D-volume $V$ and "$|$" denotes the covariant derivatives with respect to the Riemann manifold defined by the Levi-Civita
connections $\{^{\mu}_{\alpha\beta}\}$ and ${ W}^{\alpha}= \hat{\delta\Gamma}^{\epsilon}_{\beta\epsilon} g^{\beta\alpha}- \hat{\delta \Gamma}^{\alpha}_{\beta\gamma} g^{\beta\gamma}$. It is quite usual in the literature, in the framework of both, cosmological astrophysical scenarios, to assume that the surface integral must be neglected because the distance from any matter to $\partial V$ can be considered large enough to say that $W^{\mu}\rightarrow 0$ in such limit. However, it is possible to see that the 3D hyper-surface term could be a source for the cosmological parameter\cite{rb1,rb2}.

To describe the idea in terms of the geodesic dynamics, we can consider the geodesic equations in presence of an external force on a Riemann manifold. The equations take the form
\begin{eqnarray}\label{1.1geodesica}
\frac{\partial^{2}\,x^{\mu}}{\partial\,S^{2}}+\{^{\mu}_{\alpha\beta}\}\,U^{\alpha}U^{\beta}=f^{\mu},
\end{eqnarray}
in which $f^{\mu}$ are the components of the external force, that we consider as related with some interaction. In this work we shall deal only with electromagnetic interactions. The geodesics of the equation (\ref{1.1geodesica}) give us the path of a test particle over a gravitational background and describes the influence of an external and non-gravitational force.

Our purpose is to extend the theory in order to obtain a more general geometrical frame in an extended manifold where the new connection dynamics be absent of the external force. The new extended geodesic is
\begin{eqnarray}\label{1.2geodesica}
\frac{\partial^{2}\,x^{\mu}}{\partial\,S^{2}}+\Gamma^{\mu}_{\alpha\beta}\,U^{\alpha}U^{\beta}=0,
\end{eqnarray}
in which $\Gamma^{\mu}_{\alpha\beta}$ is a non-Riemannian connection. The Riemannian and non-Riemannian connections are related through
\begin{eqnarray}\label{1.3conerel}
\Gamma^{\mu}_{\alpha\beta}=\{^{\mu}_{\alpha\beta}\}+K^{\mu}_{\alpha\beta},
\end{eqnarray}
where $K^{\mu}_{\alpha\beta}\equiv \delta \Gamma^{\mu}_{\alpha\beta}$ are the components of the contortion tensor, which are responsible for the displacement from the Riemmann manifold [defined by the Levi-Civita connections $\{^{\mu}_{\alpha\beta}\}$], to the extended manifold defined by the connections: $\Gamma^{\mu}_{\alpha\beta}$. Notice that $\delta \Gamma^{\mu}_{\alpha\beta}$ are not variations in the sense of a series development, so that our formalism will be non-perturbative. In general, the contortion contributions are originated by the torsion and non-metricity, which are defined by \cite{05,051}:
\begin{equation}\label{a5}
K^{\mu}_{\alpha\beta}=\,-\frac{g^{\nu\mu}}{2}\{T^{\rho}_{\beta\nu}\,g_{\alpha\rho}\,
+\,T^{\rho}_{\alpha\nu}\,g_{\rho\beta}\,-\,T^{\rho}_{\beta\alpha}\,g_{\rho\nu}\,+\,N_{\alpha\nu\beta}\,+\,N_{\nu\beta\alpha}\,-\,N_{\alpha\beta\nu}\},
\end{equation}with \begin{eqnarray}\label{a6}
T^{\mu}_{\alpha\beta}&=&\Gamma^{\mu}_{\beta\alpha}-\Gamma^{\mu}_{\alpha\beta},\\
\label{a7}
N_{\alpha\beta\gamma}&=& g_{\beta\gamma ;\alpha},\end{eqnarray}
for a coordinate basis. In order to obtain a path which is coherent with the background source and the interaction, we must choose non-Riemannian connections according to
\begin{eqnarray}\label{1.4contorinter}
K^{\mu}_{\alpha\beta}\,U^{\alpha}U^{\beta}=-f^{\mu}.
\end{eqnarray}
We could suppose that we are dealing with observers such that $|U|^{2}=1$, so that
\begin{eqnarray}\label{1.5contor}K^{\mu}_{\alpha\beta}=-f^{\mu}\,U_{\alpha}U_{\beta}.
\end{eqnarray}
We must remark that the equation (\ref{1.5contor}) is not the more general prescription, but the simplest one in which new connection is torsion free, and all the non-Riemannian contributions are related to the non-metricity. The non-metricity takes the form
\begin{eqnarray}\label{1.6nome}
g_{\nu\pi\,;\rho}&=&g_{\nu\pi\,|\rho}-K^{\mu}_{\nu\rho}\,g_{\mu\pi}-K^{\mu}_{\pi\rho}\,g_{\nu\mu} \nonumber \\
&=&\left(f_{\pi}U_{\nu}+f_{\nu}U_{\pi}\right)U_{\rho},
\end{eqnarray}
where we have used the expression (\ref{1.5contor}). The presence of non-metricity is very important, because can be the reason why the theory is not integrable [see Appendix A].

\section{Einstein equations from a non-Riemannian geometry}\label{eins}

In this work we shall study the possible origin of the cosmological parameter due exclusively to the non-metricity of the extended variety. The study of surface terms in the action with the use of the specific geometry that we develop in present paper could be the object of further work. Under previous considerations we extreme the action: $\delta S
=0$, and we obtain that
\begin{equation}\label{tn4}
R_{\alpha\beta}\,-\,\frac{1}{2}\,{\cal R}\, g_{\alpha\beta}\,+\,\Lambda_1(x) \,g_{\alpha\beta}\,=\,0.
\end{equation}
Notice that all tensors are defined with respect to the non-Riemannian complete connections. The extended Einstein's equations (\ref{tn4}) are obtained in absence of matter in the frame of the non-Riemannian geometry, but could be re-interpreted as the effective Einstein's equations in the Riemannian manifold, with presence of matter:
\begin{equation}\label{tn5}
\bar{R}_{\alpha\beta}\,-\,\frac{1}{2}\,\bar{{\cal R}}\,g_{\alpha\beta}\,+\,\Lambda_1(x)\,
g_{\alpha\beta}\,=\,k\,\bar{T}_{\alpha\beta}.
\end{equation}
Here we must remark that $\bar{T}_{\alpha\beta}$ is geometrically induced by the
non-Riemannian characteristics of the manifold which has not a geometric explanation in the Riemannian frame: those observers who are building his physics with the Levy-Civita connections, of the Riemannian manifold, will  interpret $\bar{T}_{\alpha\beta}$ as source of the physical fields. Therefore $\bar{T}_{\alpha\beta}$ can be viewed as a geometrically induced energy-momentum tensor, that is the source of matter. This is due exclusively by contortion terms, $K^{\mu}_{\alpha\beta}$:
\begin{eqnarray}
k \,\bar{T}_{\alpha\beta}\,=\,-K^{\mu}_{\mu\beta |\alpha}\,+\,K^{\mu}_{\alpha\beta |\mu}
\,-\,K^{\sigma}_{\mu\beta}\,K^{\mu}_{\sigma\alpha}\,+\,K^{\sigma}_{\alpha\beta}\,K^{\mu}_{\sigma\mu}.
\label{4.5tensem}
\end{eqnarray}
The contribution of the cosmological parameter must be taken into account:
\begin{eqnarray}\label{4.5lambda1}
\Lambda_1(x)\,g_{\alpha\beta}\,\delta g^{\alpha\beta}\,&=&\,N_{\alpha\beta\mu}\,W^{\mu\alpha\beta},\end{eqnarray}
where we can distinguish between Riemannian and non-Riemannian parts of the tensor $W^{\mu\alpha\beta}\,=\,\bar{W}^{\mu\alpha\beta}\,+\,\Delta W^{\mu\alpha\beta}$. Then, is easy to see that $N_{\alpha\beta\mu}\,\Delta W^{\mu\alpha\beta}\,=\,0$, and the equation (\ref{4.5lambda1}) is reduced to $\Lambda_1(x)\,g_{\alpha\beta}\,\delta g^{\alpha\beta}\,=\,N_{\alpha\beta\mu}\,\bar{W}^{\mu\alpha\beta}$. On the other hand, we could identify another contribution to the cosmological parameter providing by an extra contribution of the terms that induce the effective energy momentum tensor:
\begin{eqnarray}\label{4.5lambda2} \Lambda_2(x)\,&=&\,\frac{1}{2}g^{\lambda\gamma}\,\left(K^{\mu}_{\mu\gamma |\lambda}\,-\,K^{\mu}_{\lambda\gamma |\mu}
\,+\,K^{\sigma}_{\mu\gamma}\,K^{\mu}_{\sigma\lambda}\,-\,K^{\sigma}_{\lambda\gamma}\,K^{\mu}_{\sigma\mu}\right) \equiv -\frac{k}{2}g^{\lambda\gamma}\,\bar{T}_{\lambda\gamma}.
\end{eqnarray}
The total cosmological parameter is $\Lambda (x)\,=\,\Lambda_1(x)\,+\,\Lambda_2(x)$. We can notice that both cosmological parameters, $\Lambda_1$ and $\Lambda_2$, are originated in the existence of a non-zero non-metricity. Therefore,
we obtain that the Einstein equations can be re-written with a variable cosmological parameter $\Lambda(x)$ in absence of physical sources
\begin{equation}\label{tn6}
\bar{R}_{\alpha\beta}\,-\,\frac{1}{2}\,\bar{{\cal R}}\,g_{\alpha\beta}\,+\,\Lambda(x)\,
g_{\alpha\beta}\,=0,
\end{equation}
where the $\Lambda_2(x)$-term, which is originated by the contortion, was absorbed by the cosmological parameter $\Lambda(x)$.

\section{Gravitational Waves}\label{secwave}

In order to describe gravitational waves\cite{gw, gw1,gw2,gw3,gw4,gw5,gw6,gw7}, we shall use the fact that the Riemannian metric tensor is $\bar{g}_{\alpha\beta}\,=\,g_{\alpha\beta}\,+\,\delta g_{\alpha\beta}$, to write \begin{eqnarray}\label{4.6}0\,=\,\bar{g}_{\alpha\beta\,|\gamma}\,=\,g_{\alpha\beta\,|\gamma}\,+\,\delta g_{\alpha\beta\,|\gamma},
\end{eqnarray}
where the covariant derivatives are done with respect to the Levi-Civita connections $\,\{^{\alpha}_{\beta\gamma}\}\,=\,\Gamma^{\alpha}_{\beta\gamma}-K^{\alpha}_{\beta\gamma}$. Therefore, we obtain that
\begin{eqnarray}
\delta g_{\alpha\beta\,|\gamma}\,=\,\frac{1}{2}\,G_{\alpha\beta\gamma\nu\sigma}\,\delta g^{\nu\sigma}
\end{eqnarray}
where we have used the fact that
\begin{equation}
G_{\alpha\beta\gamma\nu\sigma}\,=\,g_{\nu\beta}\,\left(g_{\alpha\sigma\,,\gamma}\,+\,g_{\gamma\sigma\,,\alpha}\,-\,g_{\beta\gamma\,,\sigma}\right)\,
+\,g_{\nu\alpha}\,\left(g_{\beta\sigma\,,\gamma}\,+\,g_{\gamma\sigma\,,\beta}\,-\,g_{\beta\gamma\,,\sigma}\right).
\end{equation}
Therefore, we massive equation of motion for gravitational waves
\begin{eqnarray}\label{4.6w2}
\bar{\square}\,\delta g_{\alpha\beta}\,+\,\Xi^{\mu\nu}_{\alpha\beta}\,\delta g_{\mu\nu}=0,
\end{eqnarray}
with $\Xi^{\mu\nu}_{\alpha\beta}\,=\,\frac{g^{\gamma\rho}}{2}\,\left({{G_{\alpha\beta\gamma}}^{\mu\nu}}_{|\rho}\,
+\,\frac{1}{2}\,{G_{\alpha\beta\gamma}}^{\pi\tau}\,{G_{\pi\tau\rho}}^{\mu\nu}\right)$. In order to obtain a Riemannian massless equation for gravitational waves, we need the following condition to be fulfilled
\begin{eqnarray}\label{4.6cond}
\delta g_{\alpha\beta\,;\gamma\rho}\,=\,-\delta g_{\nu\beta\,|\gamma}\,K^{\nu}_{\alpha\rho}- \delta g_{\alpha\nu\,|\gamma}\,K^{\nu}_{\beta\rho}-\delta g_{\alpha\beta\,|\nu}\,K^{\nu}_{\gamma\rho}-\left(\delta g_{\nu\beta}\,K^{\nu}_{\alpha\gamma}\,+\,\delta g_{\alpha\nu}\,K^{\nu}_{\beta\gamma}\right)_{;\rho}.
\end{eqnarray}
Once the condition (\ref{4.6cond}) is fulfilled, we obtain massless gravitational wave equation in absence of sources with respect to the Riemannian geometry:
\begin{eqnarray}\label{4.6w3}
\bar{\square} \delta g_{\alpha\beta}\,=\,0.
\end{eqnarray}
We must remark that the equation (\ref{4.6w3}) is effective on the Riemannian manifold and it is obtained from the distortion of an inner product over a given extended manifold. The equation (\ref{4.6cond}) describes a very specific condition, which is not fulfilled in most of cases and therefore, the induced effective gravitational wave equation on the Riemannian geometry can be massive in a more general case. It is important to notice that massive gravitons and the cosmological parameter, are a possible alternative to explain the origin of dark energy (see \cite{alvesmirandaaraujo}). \\

\section{Gravito-electrodynamics}\label{ge}

An interesting case, where the source of the geodesic equations (\ref{1.1geodesica}), is given by electromagnetic fields $A^{\alpha}$. In this
case the geodesic equations on the Riemannian manifold are
\begin{eqnarray}\label{4.1geodesica}
\frac{\partial^{2}\,x^{\mu}}{\partial\,S^{2}}+\{^{\mu}_{\alpha\beta}\}\,U^{\alpha}U^{\beta}=\,q\,F^{\mu\beta}\,U_{\beta},
\end{eqnarray}
where, in absence of torsion $F_{\mu\beta}=A_{\beta |\mu}-A_{\mu |\beta}$ is the electromagnetic tensor. As we shall seen early, this is the same equation that those without external force, but with connections:
\begin{eqnarray}\label{4.2cone}\Gamma^{\alpha}_{\beta\gamma}=\,\{^{\alpha}_{\beta\gamma}\}-\frac{q}{2}\left({F^{\alpha}}_{\beta}\,U_{\gamma}\,
+\,{F^{\alpha}}_{\gamma}\,
U_{\beta}\right),
\end{eqnarray}
in which $K^{\alpha}_{\beta\gamma}\,=\,-\frac{q}{2}\left({F^{\alpha}}_{\beta}\,U_{\gamma}\,+\,{F^{\alpha}}_{\gamma}\,U_{\beta}\right)$. We shall choose the simplest contortion in order to obtain a connection without torsion. In this case, the non-metricity is
\begin{eqnarray}\label{4.3nome}
g_{\alpha\beta\,;\gamma}=\,q\,{F^{\nu}}_{\gamma}\left(U_{\alpha}\,g_{\nu\beta}\,+\,U_{\beta}\,g_{\nu\alpha}\right),
\end{eqnarray}
where the integrability condition of the equation (\ref{1.7bisinte}) is fulfilled\begin{footnote}{We check integrability for the connection of Eq. (\ref{4.2cone}) by following the equation (\ref{1.7inte}), and taking into account the fact that $F\,=\,d(A)\,+\,A\,\cdot\,T$. Then, the condition of integrability must be checked by doing the following integration \begin{eqnarray}\label{footnote1}
\oint_{\partial\Omega} d|v|&=&\oint_{\partial\Omega} U\cdot V\,V^{\nu}\left[d(A)\right]_{\nu\gamma}\,dx^{\gamma}\,+\,\oint_{\partial\Omega} \,U\cdot V\,V^{\nu}\,A_{\mu}\,T^{\mu}_{\zeta\nu}\,dx^{\zeta},
\end{eqnarray}
where $\partial \Omega$ is a "loop", which is closed over the background geometry. This loop represents the closed path over that we parallel transport vector $V$, and $\Omega$ is the inner sub-manifold contained by $\partial \Omega$. In the present case the most simple option which guarantees the integrability, is the choice $T^{\mu}_{\zeta\nu}=0$. Then, by the use of Stokes theorem we obtain
\begin{eqnarray}\nonumber
\oint_{\partial\Omega} d|v|&=&\int_{\Omega}\,U\cdot V\,\left[d(V)\right]^{\nu}_{\kappa}\,\left[d(A)\right]_{\nu\zeta}\,dx^{\zeta}\wedge dx^{\kappa}=\\\nonumber&=&\int_{\Omega}U\cdot V\,\left({V^{\nu}}_{,\kappa}F_{\nu\zeta}-{V^{\nu}}_{,\zeta}F_{\nu\kappa}\right)\,dx^{\zeta}\otimes dx^{\kappa}=0.
\end{eqnarray}
The first term in the r.h.s. of the equation (\ref{footnote1}) results to be zero over a closed path. Then, by choosing $T^{\mu}_{\zeta\nu}=0$, we are adopting the simplest choice that guarantees integrability.}\end{footnote}:
\begin{eqnarray}\label{4.4inte}
\oint d|v|=\oint U\cdot V\,V^{\nu}\left[d(A)\right]_{\nu\gamma}\,dx^{\gamma}=0,
\end{eqnarray}
in which $F=d(A)$, as usual in absence of torsion.\\

By following equation (\ref{ac1}), the action (\ref{a1}) takes the form
\begin{eqnarray}\label{4.7ac}
{\cal S}\,=\,\frac{1}{2\kappa}\, \int_{V}d^{4}x\,\sqrt{-g}\,\left({\cal \bar R}\,+\,\Delta{\cal R}\right).
\end{eqnarray}
Under charge conservation, we obtain
\begin{equation}\Delta R\,=\,\frac{1}{4}F^{\mu\nu}F_{\mu\nu}\,+\left(\,\{^{\mu}_{\rho\psi}\}\,F_{\kappa\mu}\,U^{\rho}\,
+\,\{^{\mu}_{\kappa\psi}\}\,F_{\rho\mu}\,U^{\rho}\,-\,F_{\rho\kappa\,,\psi}\,U^{\rho}\,\right)g^{\kappa\psi}.\label{4.7aac}
\end{equation}
The first term in (\ref{4.7aac}) is interpreted as an effective electromagnetic lagrangian of matter for the effective Riemannian dynamics: ${\bar {\cal L}}\,=\,\frac{1}{4}F^{\mu\nu}F_{\mu\nu}\,$. The second one originates the $\Lambda_2 (x)$ contribution in the cosmological parameter. An observer which moves on a Riemannian manifold could not explain the origin of both terms. The geometrical action originated by the connection of the equaiton (\ref{4.2cone}) describes a scenario which induces the effective dynamics observed by a Riemannian observer over a background metric with external electromagnetic fields and nonzero cosmological parameter:
\begin{eqnarray}\label{4.8}
\delta {\cal S}=\int d^4x\,\sqrt{-g}\left({\bar G}_{\mu\nu}\,+\,\Lambda_1(x)\,g_{\mu\nu}\,+\,k\,{\bar T}_{\mu\nu}\right)\,\delta g^{\mu\nu}.
\end{eqnarray}
Here,  $k{\bar T}_{\mu\nu}\,=\,-2\frac{\delta \,{\bar {\cal L}}}{\delta\,g^{\mu\nu}}\,+\,g_{\mu\nu}\,{\bar {\cal L}}$ is an effective energy-momentum tensor for electromagnetic fields. On the Riemannian manifold the electromagnetic dynamics being given by the equations:
\begin{eqnarray}\label{4.9max}
d(\ast F)&=&\ast J,\\\nonumber d(F)&=&0.
\end{eqnarray}
Notice that current terms are originated in the non-Riemannian part of the connections (\ref{1.3conerel}). This is an expected behavior because such extra part is related to electrodynamics by the definition in the equation (\ref{4.2cone}). The second line of these equations tell us that magnetic monopoles cannot be present, due to the absence of torsion on the Riemman manifold.

\section{Final Comments}\label{fc}

We have considered the variation of the geometrical contribution of the EH action, taking into account the conditions of integrability in the boundary conditions. It is interesting that some physical phenomena can be described when we consider a displacement from a background Riemann manifold, such that the non-metricity on the extended manifold is nonzero. In particular, in this work we have dealt with emergent gravitational waves and electromagnetic interactions in a classical description. In the second case the absence of monopoles is compatible with the scenario studied by Ponce de Le\'on in\cite{poncedeleon}. Of course, we have not made any assumption about the metric tensor. Our results remain valid for
both, a spherically symmetric and static metrics, that enable us to use the Ponce de Le\'on results in a charged perfect fluid interpretation. Therefore, in a co-moving frame we perceive a central massive and charged object, compatible with an electric monopole surrounded by a perfect fluid that can be obtained from the effective equation (\ref{4.9max}). In the case in which our metric were not spherically symmetric and static, we would describe the electromagnetic interaction from a geometrical source in order to obtain effective gravitational waves, according to the equation (\ref{4.6w2}). In the case of gravitational waves the action of equation (\ref{a1}) is expressed over a non-Riemannian geometry implying that the equivalent action over a Riemannian geometry must have extra terms. This is equivalent to the exposed by Visser in \cite{visser98}. In our case such terms are induced from the extended geometry, which is originated in the non-metricity of the extended manifold (i.e., the covariant derivative of the tensor metric on the extended manifold is nonzero: $g_{\alpha\beta;\gamma}\neq 0$). This means that the physical interpretation of phenomena is determined by the geometrical description of reality in which the observer describes the physical system. This interpretation of the physical phenomena could be very interesting in a cosmological framework, For instance, the existence of a cosmological parameter $\Lambda(x)$ in the universe, could be interpreted as an empirical prove of a non-conservative norm of vectors and tensors (originated by a nonzero non-metricity), along the expansion of the universe without physical sources, described on a Riemannian manifold, in agreement with the equation (\ref{tn6}). Another possible interpretation is an universe with cosmological parameter $\Lambda_1(x)$, such that the expansion is produced by a physical source $\bar{T}_{\alpha\beta}$, in agreement with the equation (\ref{tn5}), explained on a Riemann manifold. However, this topic is beyond the scope of this work, and deserves a more rigorous study for a future work. \\

\section*{Acknowledgements}

\noindent The authors acknowledge CONICET, Argentina (PIP 11220150100072CO) and UNMdP (EXA852/18) for financial support.
\bigskip

\section*{Appendix A: Conditions of integrability in boundary terms}\label{int}

The condition of integrability expresses the fact that we can univocally assign a norm to any vector in a given point, which can be written as
\begin{eqnarray}\label{1.7bisinte}\oint d|V|=0.\end{eqnarray}
Non integrability is due to the fact that, when we parallel transport of a vector field along a closed path, then the "norm" could change in proportion to non-metricity according to:
\begin{eqnarray}\label{1.7inte}\oint d|V|&=&\frac{1}{2|V|}\oint g_{\nu\pi\,;\rho}\,V^{\nu}\,V^{\pi}\,dx^{\rho}=\\\nonumber
&=&\frac{1}{|V|}\oint f \cdot V\,\,U\cdot V\,\,U_{\rho}dx^{\rho}.\end{eqnarray}
It is easy to notice that orthogonality between $f$ and $V$, or between $U$ and $V$, grants the nullity of the integral in eq. (\ref{1.7inte}), but $V$ must be any vector field, and therefore we {\it cannot} say that $f\cdot V=0$ or $U \cdot V =0$. We could think that eq. (\ref{1.7inte}) is providing us with a sufficient condition over the possible velocity of observers (the field of tetra-velocity $U$ could be non-rotational), in order to obtain integrability. However, it is not necessary, as we shall see later, when we analyze gravitomagnetic interactions. A non-intuitive fact is that torsion plays a significative role over integrability. We are performing a closed integral in order to show how the norm of a vector changes, but a closed path must be affected by torsion. We could expect that a Burgers vector arises. This is a measure of how a given path, which is closed in a torsion-free geometry, fails to be closed in the torsional one. Then, an extra term appears in eq. (\ref{1.7inte}), and such term could complete the integral in order to be really closed. We say that torsion does not affect geodesics, but it affects integrability, so that the choice of an appropriate torsion could be an ingredient which results to be appropriate to make integrable the theory when were possible obtaining a torsion that compensates the integral of the r.h.s. of eq. (\ref{1.7inte}). In the example developed in Sect. (\ref{ge}), we shall show that the most simple choice for a given connection ({\it i.e.}, the null torsion choice), is a good choice (but not the unique), in which integrability is maintained [see footnote 1].

\end{document}